\documentclass[aps,10pt,prl,twocolumn,showpacs, showkeys, superscriptaddress]{revtex4}
\usepackage[intlimits]{amsmath}
\usepackage{amssymb}
\usepackage{graphicx}
\usepackage{epstopdf}

\usepackage{natbib}
\begin{document}
\title{Topological transitions in carbon nanotube networks via nanoscale confinement}

\author{Sivasubramanian Somu}
\affiliation{NSF Nanoscale Science and Engineering Center for High-rate Nanomanufacturing, Department of Mechanical and Industrial Engineering, Northeastern University, Boston MA 02115}
\author{Hailong Wang}
\affiliation{Group for Simulation and Theory of Atomic-Scale Material Phenomena ({\it st}AMP), Department of Mechanical and Industrial Engineering, Northeastern University, Boston MA 02115}
\affiliation{NSF Nanoscale Science and Engineering Center for High-rate Nanomanufacturing, Department of Mechanical and Industrial Engineering, Northeastern University, Boston MA 02115}
\author{Younglae Kim}
\affiliation{NSF Nanoscale Science and Engineering Center for High-rate Nanomanufacturing, Department of Mechanical and Industrial Engineering, Northeastern University, Boston MA 02115}
\author{Laila Jaberansari}
\affiliation{NSF Nanoscale Science and Engineering Center for High-rate Nanomanufacturing, Department of Mechanical and Industrial Engineering, Northeastern University, Boston MA 02115}
\author{Myung Gwan Hahm}
\affiliation{NSF Nanoscale Science and Engineering Center for High-rate Nanomanufacturing, Department of Mechanical and Industrial Engineering, Northeastern University, Boston MA 02115}
\author{Bo Li}
\affiliation{NSF Nanoscale Science and Engineering Center for High-rate Nanomanufacturing, Department of Mechanical and Industrial Engineering, Northeastern University, Boston MA 02115}
\author{Taehoon Kim}
\affiliation{NSF Nanoscale Science and Engineering Center for High-rate Nanomanufacturing, Department of Mechanical and Industrial Engineering, Northeastern University, Boston MA 02115}
\author{Xugang Xiong}
\affiliation{NSF Nanoscale Science and Engineering Center for High-rate Nanomanufacturing, Department of Mechanical and Industrial Engineering, Northeastern University, Boston MA 02115}
\author{Yung Joon Jung}
\affiliation{NSF Nanoscale Science and Engineering Center for High-rate Nanomanufacturing, Department of Mechanical and Industrial Engineering, Northeastern University, Boston MA 02115}
\author{Moneesh Upmanyu}
\affiliation{NSF Nanoscale Science and Engineering Center for High-rate Nanomanufacturing, Department of Mechanical and Industrial Engineering, Northeastern University, Boston MA 02115}
\affiliation{Group for Simulation and Theory of Atomic-Scale Material Phenomena ({\it st}AMP), Department of Mechanical and Industrial Engineering, Northeastern University, Boston MA 02115}
\author{Ahmed Busnaina}
\affiliation{NSF Nanoscale Science and Engineering Center for High-rate Nanomanufacturing, Department of Mechanical and Industrial Engineering, Northeastern University, Boston MA 02115}

\begin{abstract}
Efforts aimed at large-scale integration of nanoelectronic devices that exploit the superior electronic and mechanical properties of single-walled carbon nanotubes (SWCNTs) remain limited by the difficulties associated with manipulation and packaging of individual SWNTs. Alternative approaches based on ultra-thin carbon nanotube networks (CNNs) have enjoyed success of late with the realization of several scalable device applications. However, precise control over the network electronic transport is challenging due to i) an often uncontrollable interplay between network coverage and its topology and ii) the inherent electrical heterogeneity of the constituent SWNTs. In this letter, we use template-assisted fluidic assembly of SWCNT networks to explore the effect of geometric confinement on the network topology. Heterogeneous SWCNT networks dip-coated onto sub-micron wide ultra-thin polymer channels exhibit a topology that becomes increasingly aligned with decreasing channel width and thickness. Experimental scale coarse-grained computations of interacting SWCNTs show that the effect is a reflection of an aligned topology that is no longer dependent on the network density, which in turn emerges as a robust knob that can induce semiconductor-to-metallic transitions in the network response. Our study demonstrates the effectiveness of directed assembly on channels with varying degrees of confinement as a simple tool to tailor the conductance of the otherwise heterogeneous network, opening up the possibility of robust large-scale CNN-based devices.    
\end{abstract}
\maketitle

Harnessing the highly efficient electronic transport in carbon nanotubes continues to be a challenge due to difficulties in manipulating and packaging individual CNTs~\cite{cnt:Avouris:2007, cnt:JaveyDai:2003, cnt:Hersam:2008}. In instances where devices have been successfully integrated, issues related to device reliability and scalability remain largely unresolved. In contrast, top-down integration efforts centered around SWCNTs networks (CNNs) as electronic material of choice have enjoyed considerable success. Several device applications have been realized~\cite{cnt:SnowCampbell:2004, cnt:WuRinzler:2004, cnt:KangRogers:2007}, from gate-modulated thin film transistors (TFTs)~\cite{cnt:ZhouRogers:2004, cnt:KocabasRogers:2005, cnt:LeMieuxBao:2008} and RF oscillators~\cite{cnt:KocabasRogersZhang:2008} to wafer-scale integrated circuits on flexible substrates~\cite{cnt:Gruner:2006, cnt:CaoRogers:2008}. A key limitation of this approach is that current CNT synthesis routes yield a mixture of metallic and semiconducting nanotubes that cannot be easily separated~\cite{cnt:KrupkeKappes:2003, cnt:ArnoldHersam:2006, cnt:ZhangDai:2006}. This intrinsic electrical heterogeneity leads to networks with drastically different effective transport characteristics: as-synthesized networks can be metallic (m-CNNs) or semiconducting (s-CNNs). Since the latter response is desirable for gate modulation, it is the presence of active (metallic) elements that continues to be an issue.

Purification efforts targeted at increasing the fraction of semiconducting nanotubes hold promise yet they invariably compromise the scalability. Alternative strategies have revolved around suppressing the transport through active elements by eliminating the off-state conduction.  
Efforts to this end involve patterning heterogeneous networks within channels with controlled topology, by i) catalytic growth of CNTs along morphologically pattered substrates~\cite{cnt:KocabasRogers:2007}, or ii) via solution-based approaches~\cite{cnt:XiongBusnainaJung:2007, cnt:NouchiShiraishi:2008}. The former rely on chemical vapor deposition (CVD) to directly grow CNNs on prescribed substrates. Network topology can be tuned by controlling the density or distribution of the catalytic particles and the underlying substrate morphology, respectively. Random, partially aligned and fully aligned networks have been realized via these routes~\cite{cnt:KocabasRogers:2005}. However, the lack of control over the density and distribution of catalytic particles and the substrate morphology, especially at synthesis temperatures, limits the technique to low temperatures and results in network topologies which are coupled to the coverage.

\begin{figure}
\centering
\includegraphics[width=\columnwidth]{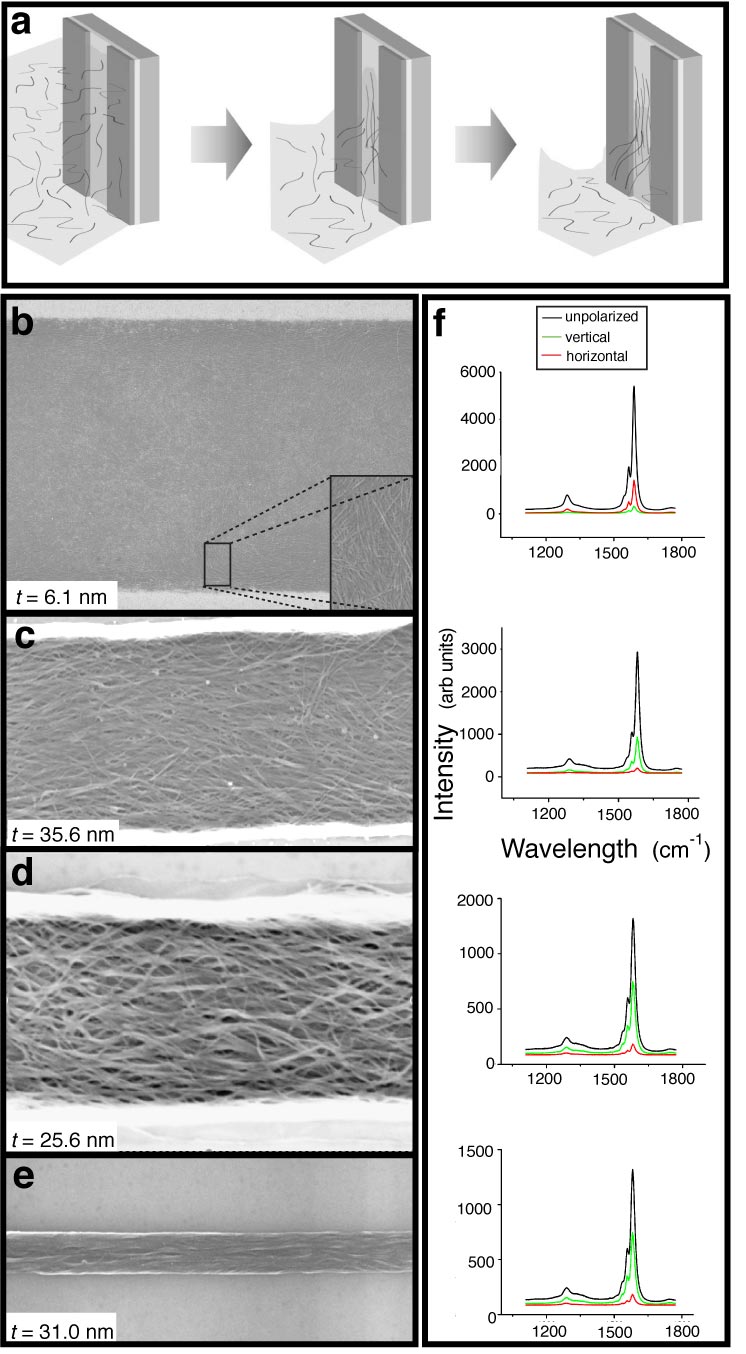}
\caption{(a) Schematic illustration of fluidic assembly of SWCNTs onto micropatterned channels consisting of alternating pattern lines of SiO$_{2}$ channels (lightly shaded) separated by photoresist (dark shaded). The receding contact line formed by the SWCNT solution following the dip-coat is exaggerated. 
(b-e) SEM micrographs of the SWCNT-patterned channels with four difference widths, (a) $W_c=9\,\mu$m, (b) $1\,\mu$m, (c) $500\,$nm and (d) $200\,$nm. The thickness $t$ of each as-deposited channel is the average of fifteen different AFM measurements (see AFM images in Supplementary Documents) and indicated on the corresponding micrograph. The thickness variation is due to that in the height of PMMA photoresists. Note that the channel width serves as the scale-bar for each micrograph. The inset in (a) is a higher magnification micrograph that shows the detail of the network topology. The corresponding polarized Raman spectra are shown below each micrograph. Vertical polarization is along the channel.\label{fig:figure1}}
\end{figure}

In this article, we use a solution-based approach to demonstrate a facile yet scalable route for synthesizing CNNs with prescribed transport characteristics. A notable aspect of this approach is the robust control over the CNN topology, which we accomplish via template-directed fluidic self-assembly of dispersions of functionalized SWCNTs onto suitably patterned channels with prescribed dimensions ($W_c\times L_c$). Plasma treated $\sim100$\,nm silicon dioxide channels are grown thermally on a silicon wafer and $150-500$\,nm  polymethylmethacrylate (PMMA) thick photoresists are used to etch nano- to micron-scale channels using e-beam lithography. The channels are then dip-coated in a dispersion solution of commercially acquired SWCNTs ($\sim0.25\%$, mean SWCNT length $\bar{l}=2.5\,\mu$m, mean SWCNT diameter $\bar{d}=1.25\,$nm) in dionized water for $5$\,min and then removed at a rate $100\,\mu$m/s. For these range of parameters, the receding contact line on the exposed channels leads to evaporation-controlled deposition of SWCNTs, as illustrated schematically in Fig.~\ref{fig:figure1}(a). For details on the synthesis technique, see Ref.~\cite{cnt:XiongBusnainaJung:2007}.

We explore the effect of nanoscale confinement on CNNs deposited on patterned channels with varying width. Figure~\ref{fig:figure1}(b)-(e) shows scanning electron micrographs (SEM) of the self-assembled SWCNTs  for four channels widths, $W_c=9\,\mu$m, $1\,\mu$m, $500\,$nm and $200\,$nm. In each case, the channel length is significantly larger ($\sim20\,\mu$m). As expected, the self-assembly of the functionalized SWCNTs is restricted to the exposed SiO$_2$ channels. Atomic force microscopy (AFM) across the channels (not shown) indicates that the assembly is uniform and consists of SWCNT multilayers with thicknesses ranging from a few to tens of SWCNT layers. The AFM images also reveal an abrupt (4-6 fold) decrease in the thickness as the channels approach nanoscale confinement, defined here as the regime for which SWCNT length becomes smaller than the channel width, $\bar{l}<W_c$. Since the SWCNT concentration in the solution and the dip-coating withdrawal rate are held constant for all templates employed in this study, the thickness change suggests that the network topology is qualitatively different. 

Direct evidence of the interplay between channel width and the CNN topology is obtained from polarized Raman spectroscopy on the as-deposited template. The spectra are depicted in Fig.~\ref{fig:figure1}(f) adjacent to the corresponding micrographs. The vertical and horizontal polarization intensity curves correspond to the SWNTs aligned along the length and width of the channels, respectively. Comparison of the Raman spectra for the different widths shows that for confined geometries, $\bar{l}<W_c$, the peak in the vertical intensity curve increases at the expense of those in the horizontal and unpolarized peaks - the SWCNTs assembly becomes increasingly aligned along the channel due to confinement. 

\begin{figure}[htb]
\centering
\includegraphics[width=0.9\columnwidth]{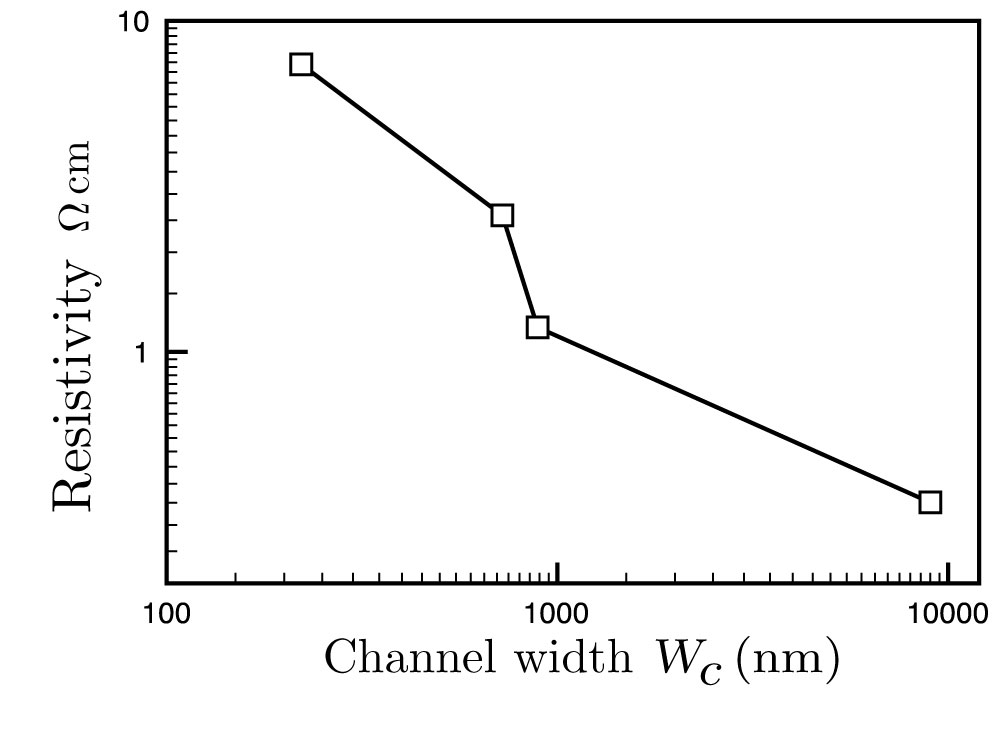}
\caption{\label{fig:figure2}
The effect of channel width on the resistivity of the assembled SWCNT networks. While the resistance is the sum of the contributions from the probe (contact) and SWCNT network, the variation is entirely a network contribution as the probe resistance is not width dependent.} 
\end{figure}

We expect the alignment to also affect electronic transport along the channels. As a first step, we have extracted the device resistivity using 3-probe resistance $R_c$ (contact + channel) measurements along the channels. To calculate the electrical resistivity $(=R_c W_ct/L_p)$, the probe length $L_p$ was obtained from SEM images at five positions along the channel. Figure~\ref{fig:figure2} shows the variation in the resistivity for the four channel widths corresponding to Fig.~\ref{fig:figure1}. As the assembled SWCNT network becomes confined, we see close to two orders of magnitude increase in the resistivity consistent with a more aligned topology that effectively shields the metallic SWCNTs within the heterogeneous network, resulting in transport that becomes increasingly controlled by CNT junction characteristics (as for example due to CNT bundling).

The assembled CNNs also vary in thickness and possibly SWCNT density, both of which are expected to modify the transport characteristics. 
To capture the effect of these network features in detail, we turn to systematic model computations of nanotube assembly on experiment-scale channels. SWCNTs coarse-grained as rigid rods are employed to gain sufficient statistics on the interplay between topology and the nature of electrical percolation within the network; the effect of channel width $W_c$ is explored using 2D computations while multi-layer, quasi-2D computations are used to study the effect of channel thickness. SWCNTs  are randomly placed onto channels with prescribed dimensions until the desired network density $\rho_{NT}$ is obtained. The stochastic nature of this procedure allows us to capture the variations in network topologies  formed due to combination of hydrodynamic forces due to the receding contact line as well as the evaporation flux that drives the fluidic assembly. The average length of the SWCNTs is matched to that in the experimental samples, $\bar{l}\sim2.5\,\mu$m and the distribution of SWCNT lengths is based on earlier experimental studies which indicate Weibull statistics~\cite{cnt:WangZhang:2006}. 

The random stick model that has been employed in earlier studies~\cite{cnt:KocabasRogers:2007} is clearly inadequate in capturing key phenomena such as CNT bundling during self-assembly (see Fig.~\ref{fig:figure2}). In order to generate realistic topologies, we further relax the network by allowing the coarse-grained SWCNTs to interact. The inter-CNT interactions, localized at the junctions, are obtained by integrating the well-known Lennard-Jones (LJ) based description of the van Der Waals between graphene surface elements. As an example, for fully aligned CNTs, the axially averaged inter-CNT interaction energy per length $U(R,r)$ is again a 6-12 LJ-type potential with constants that are scaled by surface integrals which depend entirely on the ratio of the inter-tube distance to the CNT radius $R/r$~\cite{cntr:Girifalco:2000, cntr:LiangUpmanyu:2005a}. Note that this inter-tube interaction is short-ranged and negligible for $R\ge\sqrt{3}r$, i.e. it is limited to first nearest neighbors. For partially aligned CNTs, the van Der Waals potential can again be integrated over the surfaces of the (pair) of CNTs. The resultant effective inter-CNT interactions are angular as they now depend on the degree of misalignment at the CNT-CNT junction $\phi$, i.e. $U\equiv U(R, r, \phi)$~[Wang and Upmanyu, in preparation].
In the case of fully aligned bundles, this inter-tube potential accurately describes equilibrium inter-tube spacing, cohesive energy per atom and bulk modulus.
The interactions serve as inputs for classical dynamical simulations aimed at locally relaxing the random network with respect to both translational and angular degrees of freedom of the individual SWCNTs. Sliding between the CNTs at the junctions is ignored. A time step of $1\mu$s is employed and the simulations are performed until the interaction energy associated with the network stabilizes. 
\begin{figure*}[h!tb]
\centering
\includegraphics[width=2\columnwidth]{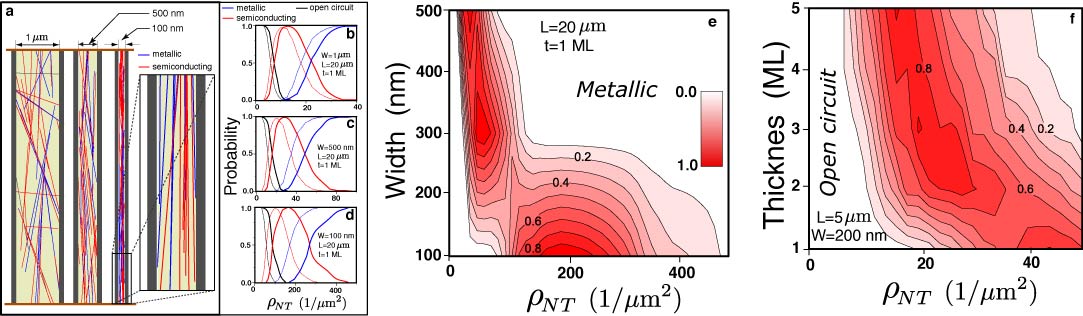}
\caption{\label{fig:figure3}
(color online): 
(a) CNNs observed in three different 2D simulations with varying channel widths, $W_c=1\,\mu$m, $500$\,nm and $100$\,nm. In each case, $5\,\mu$m length of the $20\,\mu$m long channel is shown. For clarity, detailed view of a $1\,\mu$m long segment of the $W=100$\,nm channel is also shown. Coloring scheme is based on nature of electrical transport - blue and red indicate metallic and semiconducting nanotubes, respectively. The relative number of metallic  and semiconducting nanotubes is fixed at the theoretical ratio, 1:3. (b-d) Plots of the probability of the nature of electrical percolation through the network as a function of network density, $\rho_{NT}$. The solid black, red and blue curves are probability of open circuit, semiconducting and metallic conductance across the network, respectively. The dotted lines correspond to simple stick percolation models for randomly assembled networks. (e) Contour plot of the probability of semiconducting behavior across the 2D network (shaded red) as a function of width (y-axis) and density (x-axis). The length of the channel is fixed at $L=20\,\mu$m. 
(f) Same as in (e) but for multi-layer, quasi-2D simulations with varying channel thicknesses, $t=1-5$ monolayers. The length and width of the channels are fixed, $L=5\,\mu$m, $W=200\,$nm.}
\end{figure*}

Figure~\ref{fig:figure3}(a) shows specific instances of the electrically heterogeneous network topologies obtained in three simulations with varying widths, $W_c=1\,\mu$m, 500\,nm and 100\,nm. In each case, the SWCNT network is confined, $\bar{l}/W_c < 1$ which forces the topology to become increasingly aligned with decreasing width. While the as-generated random topologies  are geometrically aligned along the channel, we find that the SWCNT interactions always work towards increasing the degree of alignment. This is not surprising as the orientation dependence of the interaction potential favors a nematic-like phase consisting of fully aligned nanotubes~\cite{lc:IslamYodh:2004}. For each relaxed CNN with a prescribed network density, the electrical transport characteristics are extracted by fixing the overall ratio of semiconducting to metallic SWCNTs to the theoretical heterogeneous density (3:1). To this end, each SWCNT is randomly assigned a metallic or semiconducting character and the overall percolation across the CNN is measured. The overall percolation can result in i) open circuit (OC), ii) semiconducting, or iii) metallic conductance across the network. Multiple simulations ($\sim100$) are performed for each channel geometry and density to determine the form of percolation. 

Figure~\ref{fig:figure3}(b)-(d) shows the percolation probability through monolayered CNNs as a function of the network density for the three widths shown in Fig.~\ref{fig:figure3}(a). All CNNs exhibits two transitions as the network density is increased: OC-to-semiconducting at low densities, and a semiconducting-to-metallic at high densities. Qualitatively similar transitions are also observed for unrelaxed networks that are used as input in random-stick models (shown as dotted lines) and comparisons with percolation in relaxed structures indicates that the enhanced alignment driven by the SWCNT interactions shifts the transitions to higher network densities. The effect of decreasing width is similar as it also enhances the alignment along the channel - for a given network density, the networks show a marked reduction in metallic transport as the alignment effectively shields the active components. This interplay between channel width and network density is more clearly seen in Fig.~\ref{fig:figure3}e, a contour plot of the probability of semiconducting percolation across the channel as a function of these two variables. The network density ranges for semiconduction increases substantially as the width is decreased below $W_c\sim300$\,nm. Note that the probability of an open circuit also increases with decreasing width, but the extent of this effect is smaller than the enhancement in semiconduction. 

The thickness of the CNNs permits control over out-of-plane confinement of the SWCNTs and we explore this effect by performing percolation studies on relaxed multilayered networks with fixed width and length. For computational efficiency, we have chosen networks with smaller lengths ($L_c=5\mu$m) and channel width ($W_c=200$\,nm). Note that the network is already confined at this width. Figure~\ref{fig:figure3} shows the contour plot associated with probability of semiconduction along the channel as a function of number of monolayers (ML), $t=1-5$ ML.  Our results show that thin CNNs have a higher probability for semiconduction and the probability decreases (at the expense of metallic behavior) rapidly within the first few monolayers. While we have not studied thicknesses greater than $t=5$ML, the contour plot shows that the marginal decrease should be much smaller.

In summary, this article demonstrates a scalable solution-based route for large-scale integrated circuits based on directed assembly of CNNs. Our results show the effectiveness of geometrical confinement induced by controlling the channel geometries, mainly their width and depth. Small, nanoscale geometries suppress the effect of active elements inherent in heterogeneous SWCNT solutions, and the challenge of large-scale integration of CNN-based devices reduces to scalable, high-fidelity lithography of assembly patterns.


This work was supported by the National Science Foundation Nanoscale Science and Engineering Center (NSEC) for High-rate Nanomanufacturing (NSF grant- 0425826). 
The experiments were conducted at the George J. Kostas Nanoscale Technology and Manufacturing Research Center at Northeastern University. In addition, YJJ acknowledges support from NSF-CMMI 0927088 and the Fundamental R\&D Program for Core Technology of Materials funded by the Ministry of Knowledge Economy, Republic of Korea. HW and MU are also grateful for support from Structural Metallics Program, ONR (N00014-06-1-0207) and the DOE-sponsored Computational Materials Science Network (CMSN).

\end{document}